# GEOSPATIAL MONITORING OF INFECTIOUS DISEASES BY UNMANNED AERIAL VEHICLES

*Chiranjib Patra*[1]
[1]Gheorghe Asachi Technical University, Iasi, Romania

## Abstract

Recent development in unmanned UAV technology paved the way for numerous applications in diverse cross discipline fields. One of the main feature of UAV s are their portability in terms of size that allows them to navigate through fairly hostile environments and collect data . This data collection leads to the interpretation of the behavior and predictability according to the analysis as presented by data science. The application of UAV to monitor the population and climate geography is well documented. But the usage of UAV to study the germs in the atmosphere is not well documented or absent. As air remains one of main medium of transmission of germs so there must be some kind of signature specific for a particular kind of germ. Using this as cue in this present communication a hypothetical model to study the spread of disease is presented. This model can help the epidemiologists to understand the mechanism of microbial traffic like for example flu getting transferred within the same species or cross species ,spatial diffusion like for example human traveling pattern and newly recognized diseases for example various type of flu and vector borne diseases like malaria , dengue etc. This model also covers some relevant scenarios like global climate change, political ecologic emergences of aerial transmitted diseases.

As by principle of disease ecology – geography the population and the climate geography are in dynamic equilibrium. Any unbalance or stressed out condition leads to the appearance of disease. These stressed out condition may be the result due to major change in the land use increased population and other geocentric factors dependent on the area under consideration. This fragile equilibrium when disrupted by the above mentioned stressors lead to the significant maladaptation, which in turn manifests into the diffusion of new diseases. In this kind of human-environment relationship, the UAV application find its pathways into the understanding of the mechanics. So by using UAV we can cover the gaps of the satellite data like long repeat times, cloud contamination low spatial resolution etc. But however the practicality of using UAV in understanding the spread of human diseases through this model fill the niche but do not replace the existing technologies.

## Introduction

Drones as of decade ago was relative expensive toy which was generally used by children for amusement. Of recently due to the adventive technologies this toy turned into one of the most useful and reliable handy surveying equipment. This drones are sometimes referred as Unmanned Ariel Vehicles (UAV) due their remotely controlled nature.

The UAV phenomenon started with the launch of Phantom drone way back in December 2012.The configuration of this drone was readily responded by the manufacturers and application developers in the diverse domain of agriculture and wildlife conservation. As the time went by the development stride took on to the current specs of the drone as pre-equipped pristine HD cameras, motorised gimbals, and many other highly intelligent features becoming the norm intoday's consumer drones. Nowadays UAV seem to capture the limelight of future delivery managers in terms of data acquisitions from various domains .It is interesting to note that the present configuration of drone the Phantom 4 is lighter, faster and longer enabling us to capture some of the highly mobile objects of interest. The Phantom 4 comes with obstacle avoiding systems, VR goggles and 360 degrees video and vision positioning systems.

This newly equipped drone can be used in airborne systems and ground based survey techniques.This flight of the drone helps to get high resolution data in fairly cheap price and is very reliable.Hence drones seem to fill the gap between satellite remote sensing and the interpretationOne of uses of UAV nowadays is to collect data related to the public health . The environmental data is one of the relevant data that affects the human health. The levels of various pollutants like CO, $CO_2$, $SO_2$, Benzene etc gives the livability index of a particular city. On board monitoring equipment has been fitted to drones to measure levels of environmental toxins and pollutants [4], [5]. Furthermore in the study of infectious diseases , UAVs provide a new pathways to collect detailed geo referenced information on environmental and other spatial variables influencing the transmission of infectious diseases. A well-document study reveals that due the changing land use , the breeding medium of these vectorized insects had changed from the forest-breeding to human population -breeding thus affecting their life cycle .These precariously changing pattern in breeding leads to an unlikely adaption called Mal-adaption. For example in south east Asia it has been experienced that the clearing of the tropical forests has lead to the changes in the anopheles mosquito life cycle. This change has lead to the increased cases of malaria in the identified area .So in the context of studying anthropogenic environmental changes can modify the transmission of zoonotic and vector-borne diseases by disrupting existing ecosystems and altering the geographic spread of human populations, animal reservoirs, and vector species .Understanding rapidly changing patterns of human settlement and vector distribution in this context is vital for predicting disease risks and effectively targeting disease-control measures.

"Coughing, sneezing, talking, bed-making, turning pages of books, etc. all generate microbial aerosols which are carried and dispersed by air movements. Inhalation of these aerosols cause allergic responses but whether or not infectious disease ensues depends in part on the viability and infectivity of the inhaled microbes and their landing sites. Desiccation is experienced by all airborne microbes; gram-negative bacteria and lipid-containing viruses demonstrate phase changes in their outer phospholipid bilayer membranes owing to concomitant changes in water content and/or temperature"[8] accurately describes how the airborne microbes work. Just in other case microbial infections like influenza poses serious threat to public health. This particular disease is highly contagious and rapidly spreading due to recent pandemics. The mode of influenza transmission events occurs via aerosol virus system in the air. These aerosol droplets are self-contained and have diameters less than 5 microns. These aerosols droplets are reasonably stable for time ranging to hours which makes it far reaching in terms of infecting the population. The main trait of these viral infection is from the respiratory system , which is basically deposited in the lower airways during inhalation. The authors in [7] describes electrostatic precipitation (ESP)-based bioaerosol sampler that is coupled with downstream quantitative polymerase chain reaction (qPCR) analysis which is ideal for offline analysis of the sample collected by UAV from various regions of identified geographical area. Similarly in another study authors [9] have developed a method called molecular and infectivity assays which can effectively determined using a fluorescent focus assay and influenza virus nucleic acid
(originating from viable and non-viable viruses) was measured using quantitative Polymerase chain reaction. In another effective low cost way of determining the presence of microbes in the aerosols have been studied by [11].The detection is based Near Infra Red Spectroscopy principle[10][11].But still the challenge lies in the development of light weight aerosol analyser that can be fitted on board of such UAV as described in Lab-O-Drone[1]

**The Work Flow Model**
The above model is simple linear model but have lot of pointers towards the emerging new technologies which are used not on very large scale. This model is applicable to the air borne diseases and vector borne diseases. The below is the simple work flow of the model discussed. The following steps of the work flow is
0. Stage-0: This step is one of the most important where the population responds to the ongoing crisis . By using the respective api of the social networks the data analytics can be made to quantify the gravity of feeds .This type of approach was used to study the *twitter* feed activity in response to a 5.8 magnitude earthquake which occurred on the East Coast of the United States (US) on August 23, 2011.[12]
1.Stage-1: (Satellite Remote Sensing Data)This data is obtained by existing technology to obtain the imagery of the land that is taken under consideration. As the resolution of the image is increased the cost is also increased . In the remote sensing data produced by the satellite has three components like spectral, spatial and temporal .Out of these three components spectral resolution is responsible for clarity of the image. Using the best known sensors on the satellite the error is around 1m approx. So this stage is only useful to collect the data related to next stage.
2. Stage-2:(Identification of 1.country 2. vegetation 3.population density 4.elevation from sea level ) This data is used to consider the set of diseases known to the environmental settings of the geographical location. Like for example tropical countries are heavily prone to malaria type of diseases than the places with extreme conditions. If there are diseases reported outside this set ,then special consideration should be given on the understanding of the particular disease in that area.
3.Stage-3:(Drone for initial Survey for domestic animals, population density, birds, flying insects) This stage is carried out if there is a need for better analysis of the particular geographical location .The movement of the cattle and the birds are important in understanding the spread of the airborne as well as vector borne disease. The cattle and birds are known for their capacity to carry vectored insects and their larvae from one place to another in a short distance. This stage can also be used to find the concentration of vectored insects in a particular area .
4. Stage-4:(Identification of houses known for any airborne or vector Borne disease if any) This is an optional stage for first time operators of the work flow. But from recent past records, the dwelling of the sick person can be identified. The statistics of the air analysis and the vectored insect population is important for the disease management to consider preventive steps.
5.Stage -5:(Mapping of the infectious areas) This stage coarsely classifies the subtle regions susceptible to the disease .

6. Stage-6:(Lab-O-Drone launch) This Lab-O-Drone [1] is an mobile smartphone based quardcopter helping in the diagnostic services in the event of ebola breakout. But this quardcopter can be used to collect sample water from shallow locations around the designated area. This sample water can be tested on board for the presence of larvae of Plasmodium knowlesi etc. Again for checking the virus in the airborne aerosols can be done from the on board low cost NIR Spectroscopy or Surface-Enhanced Raman Scattering [2][3].

7. Stage-7:(Check the air and suspended Aerosols for contamination And locate the presence of vector insects /larvae ) This is actually the dataset that can be obtained as the result of the drone flight to the designated areas.

8. Stage-8: This stage is the output of the analytics provided from the previous stage. Moreover this stage can be augmented by a system unlike UAVTechLab [6] uses the artificial intelligence paradigm to scan designated areas and try to identify injured civilians and provide the medical support. This same artificial intelligence algorithms can extended for the purpose of identifying the infected areas and provide the basic necessary medical support.

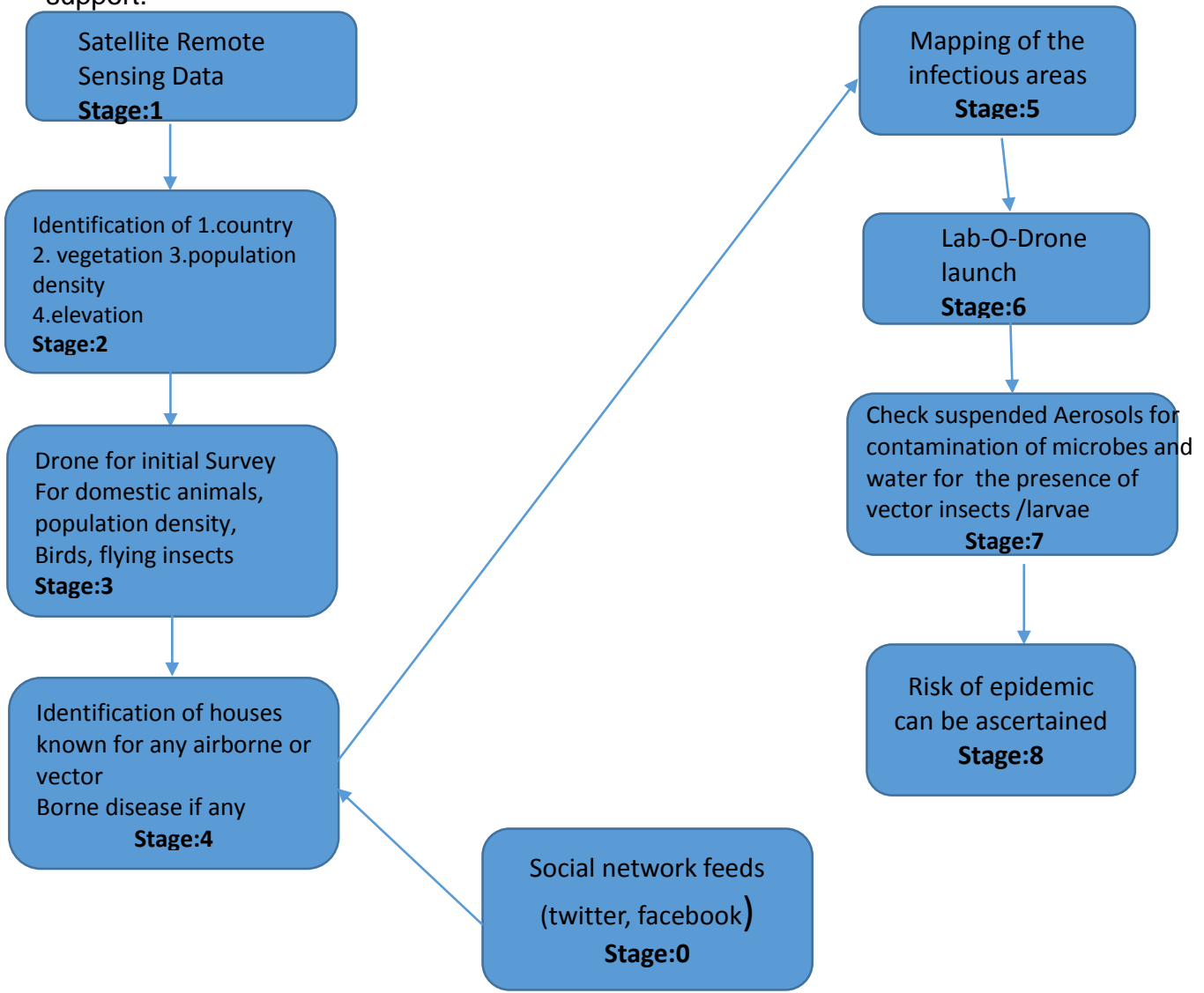

## Conclusions and Future Work

This particular model work flow will have huge impact on the infectious disease control and mitigation system with usage of UAVs. This kind of workflow model is equally applicable to rural as well as urban areas of the geographical area.This particular model also simulates the need of low cost and portable technologies for accurate on board analysis of the aerosols considered. Otherwise the UAV have to frequent flights for the collection of samples. Moreover certain improvement in flight capacity of drones has to be incorporated to collect samples from shallow water bodies.

However the implementation of this model workflow will generate important data about the insights of the geography of the diseases ,spreading dynamics both static as well as dynamic human populations. There can also be a study on effect of climate change on the geography of diseases , understanding the effect of various stressors on the human -environment system, effect of pollution on the diseases spread and also the understanding of the mechanics of the emergence of new diseases to name a few .

NB: This paper was presented at GeoMundus 2017 (http://www.geomundus.org/2017/) and was one of the winners of Travel Grant for the presentation of Abstract at the Institute for GeoInformatics , Munster, Germany